\documentclass[prl,twocolumn,floatfix,showpacs,preprintnumbers,superscriptaddress,amsmath,amssymb]{revtex4}
\usepackage[dvips]{graphics}

\def\pfhv{\hat{\mbox{\boldmath$p$}}_{\!\!{\scriptscriptstyle F}}}
\def\vp{{\bf p}}
\def\vv{{\bf v}}
\def\vy{{\bf y}}
\def\vz{{\bf z}}
\def\vH{{\bf H}}

\usepackage{graphicx}

\begin{document}

\title{Shot noise and multiple Andreev reflections in d-wave superconductors}

\author {J.C.~Cuevas}
\affiliation{Institut f\"ur theoretische Festk\"orperphysik, Universit\"at Karlsruhe, 
76128 Karlsruhe, Germany}

\author {M.~Fogelstr\"om} 
\affiliation{Institute of Theoretical Physics, Chalmers University of
Technology and G\"oteborgs University, S-41296 G\"oteborg, Sweden}

\date{\today}

\begin{abstract}
We present a theoretical analysis of the shot noise in d-wave/d-wave contacts 
with arbitrary transparency, including the contribution of multiple Andreev
reflections. The multiple charge quanta transferred in these processes are
revealed as a huge enhancement of the noise-current ratio at low voltages,
which survives for all crystal misorientations. We also show how different 
ingredients like non-magnetic impurities or a magnetic field produce very 
characteristic hallmarks in the shot noise, which can be used as a further 
test of the d-wave scenario in superconducting cuprates.
\end{abstract}

\pacs{74.50.+r, 72.70.+m, 74.72.-h, 74.80.Fp} 

\maketitle

In the last years the intensive study of the nonequilibrium current fluctuations,
known as shot noise, has provided a deeper understanding of the electronic transport 
in many different contexts \cite{Blanter2000}. The shot noise reveals aspects hidden 
in the usual conductance measurements like the statistics and charge of the carriers, 
relevant energy scales or transmission information. In the case of superconducting 
contacts, the noise has been mainly used for the analysis of the effective charges 
transferred in the different tunneling processes. This effective charge can be roughly 
defined as the noise-current ratio. A striking example is the recent observation in 
superconducting atomic-size contacts of effective charges much larger than unity 
attributed to multiple Andreev reflections \cite{Cron2001}, and in quantitative 
agreement with the theoretical expectations \cite{Cuevas1999}. Unfortunately, the 
analysis of shot noise has been mainly restricted to conventional superconductors, 
and only a few theoretical works have recently addressed noise in NIS-junctions with 
d-wave superconductors \cite{Zhu1999}. 

On the other hand, the origin and nature of the high temperature superconductivity  
in the cuprates is still an open problem. Different phase-sensitive experiments have 
provided strong indications that the order parameter in these materials has a 
dominant $d_{x^2 - y^2}$ component \cite{Harlingen1995,Tsuei2000}. However, these 
experiments have not definitively closed the debate about basic questions like the 
universality of this symmetry or the existence of subdominant components 
\cite{Covington1997,Neils2002}. In this sense, it is highly desirable to provide 
new tools which can further test the different scenarios. Thus, a natural question 
is: what can the shot noise teach us in unconventional superconductivity? For 
instance, the shot noise provides fundamental information on the charge of the 
carriers, as it was shown in complex situations like in the fractional quantum 
Hall effect \cite{Saminadayar1997} or in superconducting point contacts 
\cite{Cron2001}. On the other hand, the shot noise in a junction depends in a 
different way on the interface properties as compared with the current. In this 
sense, the study of this quantity can used as a cross-check for the different 
transport theories, and in turn it can be very valuable to solve the lack of 
consensus in the interpretation of the tunneling experiments in cuprate junctions 
\cite{Kashiwaya2000}.

In this paper, we present the first theoretical analysis of the shot noise in 
d-wave/d-wave SIS junctions of arbitrary transparency. We shall show that the
zero-frequency noise, $S$, may by large exceed the Poisson value $2eI$, where 
$I$ is the current, due to the occurrence of multiple Andreev reflections. 
In particular, at high transparencies the effective charge, $q$, defined as $q = 
S/2I$, exhibits a huge enhancement at low voltages ($q/e \gg 1$), which survives 
for all crystal misorientations. At low transparencies, contrary to the s-wave case, 
$q$ is not quantized in units of the electron charge due to the averaging over the 
anisotropic gap. We shall also show that elastic scattering mechanisms like bulk 
impurities may result in a strong reduction of the effective charge. Finally, we 
shall show how the Doppler shift of the Andreev bound states in the presence of a 
magnetic field is revealed in the shot noise. All these features are very 
characteristic of the d-wave symmetry and can be used as additional tests of
this scenario in cuprates. 

Our goal is to extend the theory of the shot noise to the case of superconducting 
cuprates. For this purpose, we consider a voltage biased contact, consisting of 
two $d_{x^2 - y^2}$ superconductors separated by a single interface of arbitrary 
transparency. The order parameter on side $i$, $i=L,R$, is rotated by $\alpha_i$ 
with respect to the surface normal, and we denote junction type by the relative 
crystal orientations as $d_{\alpha_L}$-$d_{\alpha_R}$. There are several 
experimental realizations of this system, among which the bicrystal grain-boundary 
junctions are ideal examples \cite{Alff1998}. To calculate the noise we use the 
formalism developed in Ref.~\cite{Cuevas2001}. In that work we introduced 
a formulation of boundary conditions that mimics interfaces for the quasiclassical
theory of superconductivity and that are suitable for arbitrary transparency, and
we established the machinery to determine the current fluctuations in unconventional 
junctions. Here we consider the case of point-contact-like geometry and assume
that the voltage drop takes place at the interface. Thus, to compute the noise we
first determine self-consistently the local electronic properties of the isolated 
electrodes. This includes effects on the order parameter profile and on the local 
density of states (DOS) by pair breaking caused both by quasiparticle scattering
off the interface and off homogeneously distributed impurities in the crystals
\cite{Matsumoto1995,Buchholtz1995,Fogelstrom1997,Barash1997,Poenicke1999,Poenicke2000}.
Finally, the noise is calculated using the local surface Green's and solving the 
appropriate boundary conditions for a point contact, as detailed in 
Ref.~\cite{Cuevas2001}.

The noise spectral density $S(\omega)$ is defined as

\begin{equation}
S(\omega)  =  \int d (t^{\prime}-t) \; e^{i\omega \left(t^{\prime}-t\right)} 
\langle \delta \hat{I}(t^{\prime}) \delta \hat{I}(t) + 
\delta \hat{I}(t) \delta \hat{I}(t^{\prime}) \rangle , \nonumber
\end{equation}

\noindent
where $\delta \hat{I}(t)= \hat{I}(t) - \langle \hat{I}(t) \rangle$ are the 
fluctuations in the current. We only consider the zero-frequency limit at zero 
temperature. In the case of a constant bias voltage, $V$, one can show (see 
Ref.~\cite{Cuevas2001}) that the noise oscillates in time with all the 
harmonics of the Josephson frequency, i.e. $S(t) = \sum_m S_m e^{im \phi(t)}$, 
where $\phi(t)=\phi_0+(2eV/\hbar)t$ is the time-dependent superconducting phase 
difference. We shall only consider the dc noise, denoted from now on as $S$.
Furthermore, we assume that the interface conserves the momentum of the quasiclassical 
trajectories, which allows us to write the noise as a sum over independent trajectory 
contributions: $S = \frac{1}{2} \int^{\pi/2}_{-\pi/2} d\pfhv \; S(\pfhv) \cos(\pfhv)$, 
where $\pfhv$ defines the Fermi surface position. For the angular dependence of the 
transmission coefficient we use the expression $D(\pfhv) = D \cos^2(\pfhv)/ 
[1-D\sin^2(\pfhv)]$, resulting from a $\delta$-like potential. Here $D$ is the 
transmission for the trajectory perpendicular to the interface. In the tunneling
regime one can easily demonstrate that the zero-frequency noise reaches the Poisson
value, i.e. $S=2eI$. Thus, in this limit the noise does not contain new information
as compared with the current. For this reason, we shall investigate the case of not 
too low interface transparency, $D\ge 0.1$, in which the multiple Andreev reflections 
(MAR) play a fundamental role in the transport \cite{Lofwander2001,Poenicke2002}.

\begin{figure}[h]
\begin{center}
\includegraphics[width=\columnwidth,clip=]{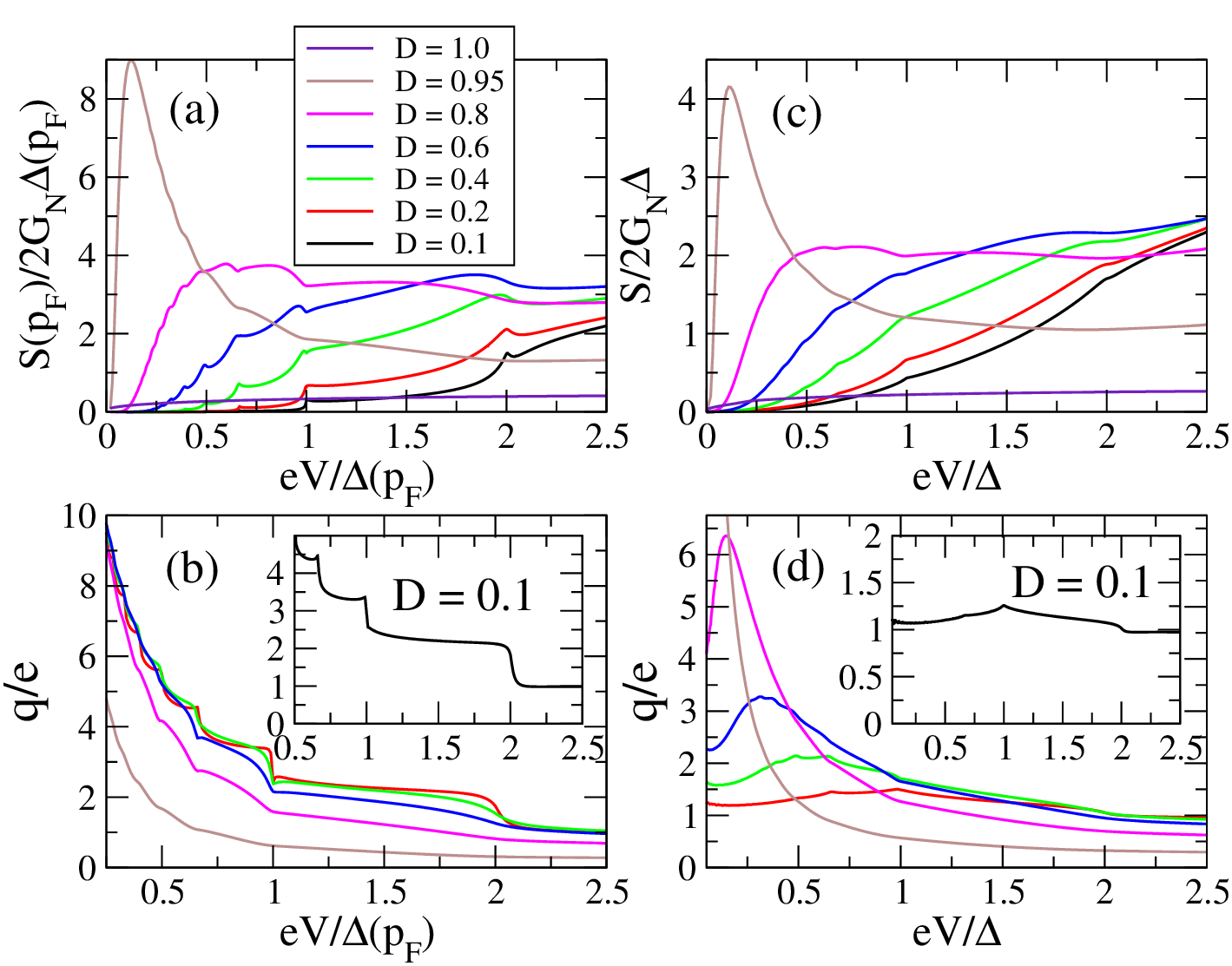}
\vspace*{-0.3truecm}
\caption{$d_0$-$d_0$ contact in the clean case: (a) Angle-resolved shot noise as 
a function of the voltage in units of the trajectory gap for different transmissions. 
$G_N$ is the normal state conductance. (b) Angle-resolved effective charge, $q=S/2I$. 
(c) Angle-averaged shot noise. The voltage is normalized by the maximum gap $\Delta$. 
(d) Angle-averaged effective charge.}
\label{0-0}
\end{center}
\end{figure}

Let us start by analyzing the case of a symmetric $d_0$-$d_0$ junction in the clean
limit. In this case, the order parameter is constant up to the surface, and there
are no bound states for any trajectory. The noise-voltage characteristics for a 
single trajectory, $S(\pfhv)$, coincide with those of isotropic s-wave superconductors
\cite{Cuevas1999}, and can be seen in Fig~\ref{0-0}(a). As a consequence of the 
occurrence of MARs, the trajectory-resolved shot noise exhibits the following 
remarkable features: (i) the presence of a pronounced subharmonic gap structure (SGS)
at voltages $eV=2\Delta(\pfhv)/n$, (ii) the noise greatly exceeds the Poisson value 
$2eI$ in the subgap region, as can be seen in Fig~\ref{0-0}(b), and (iii) in the 
tunneling regime the effective charge is quantized in units of the electron charge. 
This last feature, illustrated in the inset of Fig~\ref{0-0}(b), was used to suggest 
that the noise provides a way of measuring the charge of individual MARs in s-wave 
superconductors \cite{Cuevas1999}. Indeed, as mentioned in the introduction, these 
noise-voltage characteristics have been quantitatively confirmed in the context of 
superconducting atomic contacts \cite{Cron2001}. The natural questions now are: do 
these features survive after doing the average over the different directions in the 
Fermi surface? Can we still identify the charge of individual MARs in a d-wave 
junction?. The answers to these questions can be seen in Fig.~\ref{0-0}(c-d). First 
of all, notice that the SGS is still visible, but it is more rounded than in the 
s-wave case. It is worth remarking that it is the bulk maximum gap what is revealed 
in the SGS. Notice also that the effective charge does not show any sign of 
quantization even at low transparencies (see inset of Fig.~\ref{0-0}(d)). This is due
to the fact that different MARs contribute simultaneously for different trajectories,
and then the discreteness of $q$ is washed out. Anyway, the dominant contribution 
of MAR at high transmission is still manifested as a huge enhancement of the 
effective charge at low bias ($q \gg e$).

\begin{figure}[h]
\begin{center}
\includegraphics[width=0.85\columnwidth,clip=]{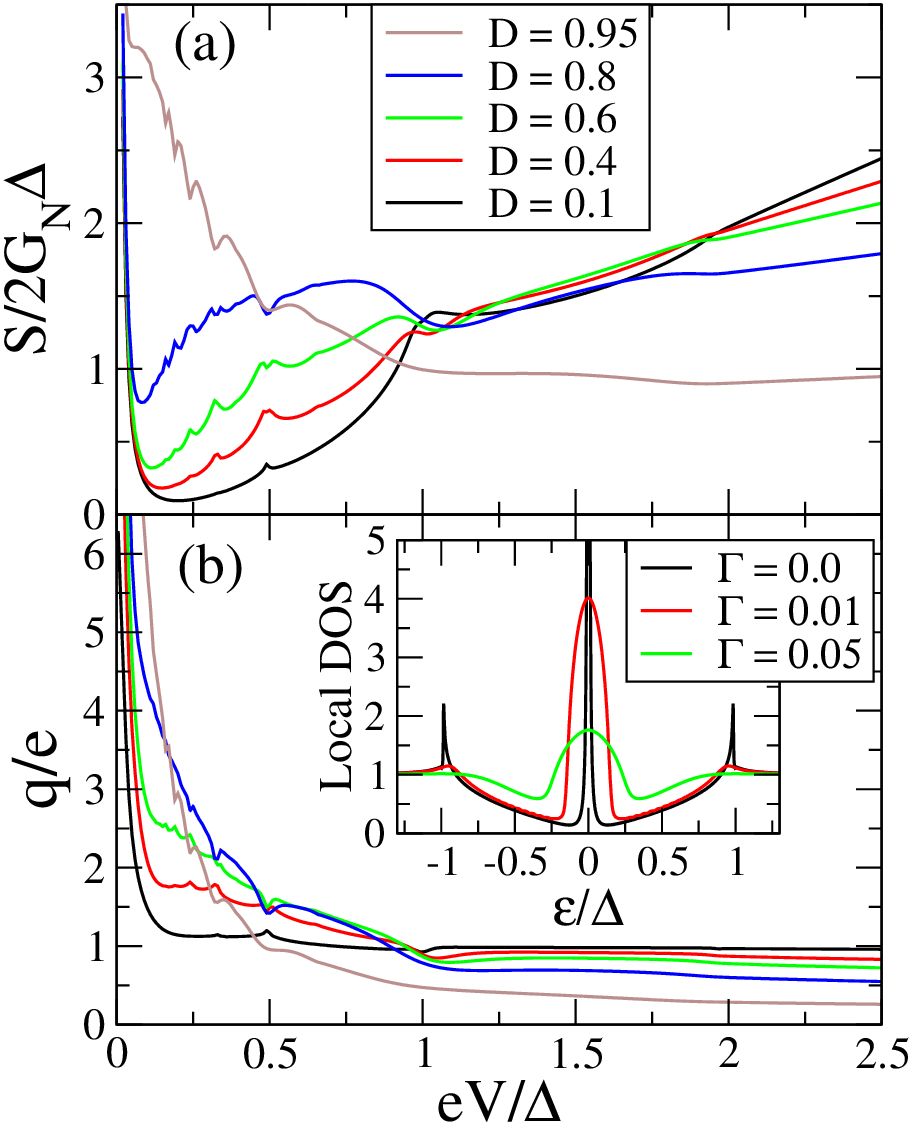}
\vspace*{-0.3truecm}
\caption{$d_{\pi/4}$-$d_{-\pi/4}$ contact in the clean case: (a) Shot noise as
a function of voltage for different transmissions. (b) Effective charge vs voltage.
The curves were computed using a small inelastic broadening ($\sim 0.003\Delta$).
Inset: local DOS at the interface for a $45^o$ misorientation for different values 
of the bulk-impurity scattering rate $\Gamma$ (Born scatterers), measured in units 
of $2\pi T_C$, where $T_C$ is the critical temperature in the clean case. $\Delta$ 
is the maximum bulk gap for the clean superconductor.}
\label{45-45}
\end{center}
\end{figure}
\vspace{-4mm}

Let us now consider the case of a $d_{\pi/4}$-$d_{-\pi/4}$ junction. In this case,
assuming specular quasiparticle scattering at the interface, an Andreev bound state
forms at zero energy for every trajectory \cite{Hu1994}. This implies that the surface
acts as a pair-breaker \cite{Matsumoto1995,Buchholtz1995} and the gap is depressed in 
the vicinity of the interface, vanishing exactly at the barrier. This order-parameter 
profile induces not only the appearance of bound states at zero energy, but also at the 
gap edges, as can be seen in the inset of Fig.~\ref{45-45}(b). As a consequence of this 
local density of states the noise exhibits a pronounced SGS due to resonant tunneling
through the bound states (see Fig.~\ref{45-45}(a)). As in the case of the current, 
see Ref.~\cite{Poenicke2002}, there is an even-odd effect in the SGS, in 
the sense that the even ($eV=\Delta/n$) structures are more pronounced. Its origin can 
be understood as follows. In this geometry there are two types of MARs which dominate 
the transport: (a) those which connect the bound states with the gap edges and (b) 
the usual ones connecting the gap edges. The first ones give rise to the SGS at 
$eV = \Delta/n$, while the second ones contribute to the whole series $eV = 2\Delta/n$. 
However, the bound states at the gap edges do not appear for every trajectory, which 
weakens the SGS due to these processes. On the other hand, as we show in 
Fig.~\ref{45-45}(b) the effective charge is not even quantized at low transparencies, 
again due to the average over the different trajectories. However, at high transparencies 
the dominant contribution of the MARs give rise to a huge enhancement of the effective 
charge at low bias. This is a robust feature which survives for all crystal 
misorientations, and it is an unambiguous signature of the fact that the MARs control 
the low voltage transport. Indeed, this pronounced increase of the noise-current ratio 
at low bias has been recently observed in symmetric bicrystal YBCO Josephson junction 
\cite{Constantinian2002} in, to our knowledge, the first experimental analysis of the 
shot noise in cuprate Josephson junctions. In this experiment a mean transparency
of $D \approx 0.01$ was estimated, but in our opinion this enhancement is due to MARs 
in high transparent conduction channels, probably due to the presence of pinholes like 
in the conventional SIS tunnel junctions of Dieleman \emph{et al.} \cite{Dieleman1997}.

\begin{figure}[h]
\begin{center}
\includegraphics[width=\columnwidth,clip=]{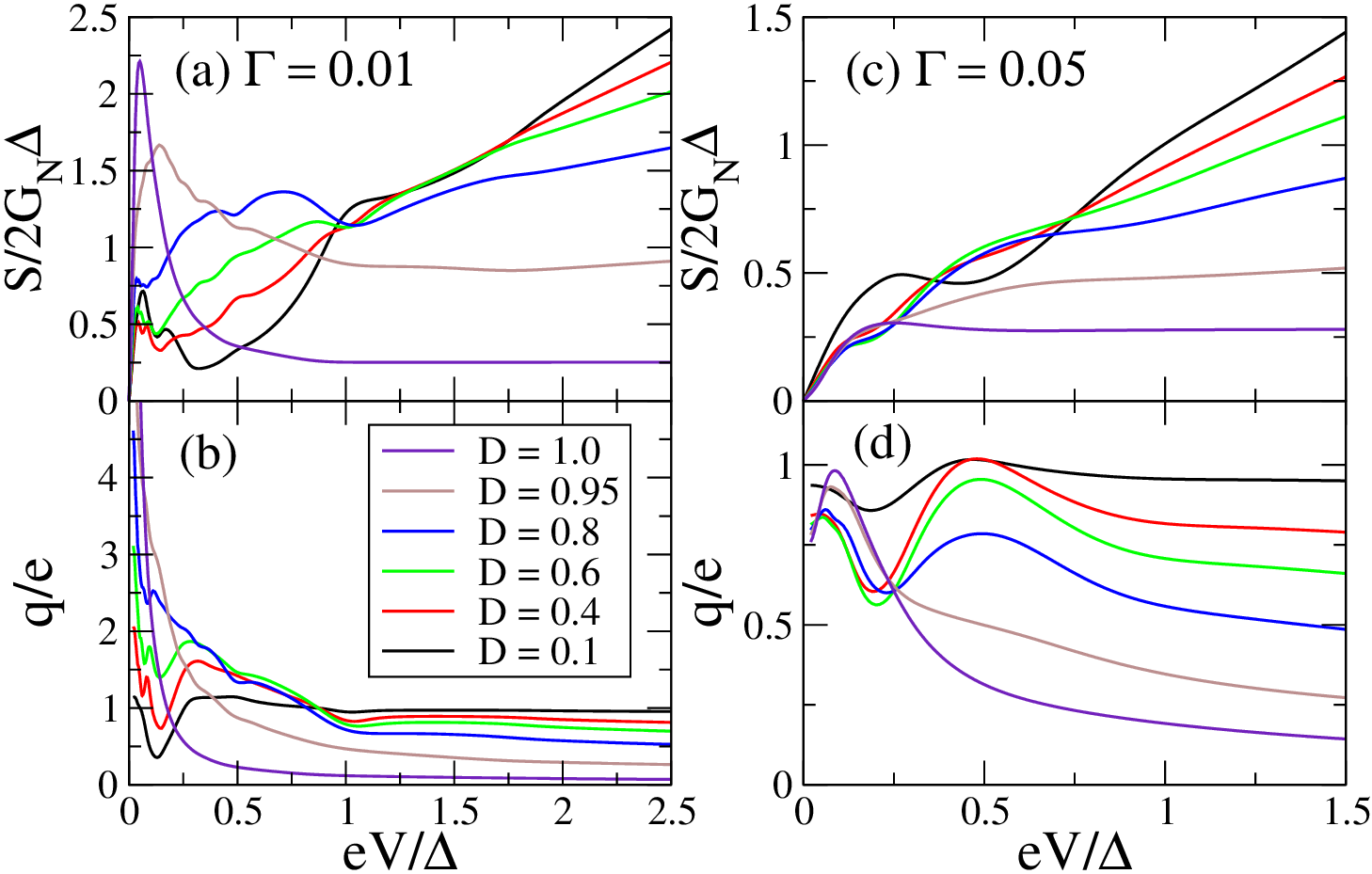}
\vspace*{-0.3truecm}
\caption{Shot noise and effective charge as a function of voltage for a 
$d_{\pi/4}$-$d_{-\pi/4}$ contact for two values of the bulk-impurity scattering rate 
$\Gamma$.} 
\label{impurities}
\end{center}
\end{figure}

In d-wave superconductors the order parameter is very sensitive to scattering from 
nonmagnetic impurities and surface roughness. In particular, it is known that these 
elastic scattering mechanisms provide an intrinsic broadening for the zero-energy 
bound states \cite{Poenicke1999,Poenicke2000}. For the case of Born scatterers this 
broadening is $\propto \sqrt{\Gamma \Delta}$, where $\Gamma=1/2\tau$ is the effective 
pair-breaking parameter locally at the surface. This is illustrated in the inset of 
Fig.~\ref{45-45}(b) for the case of bulk impurities. The interesting question now is: 
what is the signature of impurities in the shot noise of a d-wave junction? In 
Fig.~\ref{impurities} we show the shot noise and effective charge for a 
$d_{\pi/4}$-$d_{-\pi/4}$ junction for different values of the bulk-impurity scattering 
rate. As the scattering rate increases, there are two major effects that one should 
notice: (i) the disappearance of the SGS in the noise, and (ii) a reduction of the 
effective charge, specially pronounced at low voltages. Both features can be 
understood as follows: the increase of density of states in the gap region enhances 
the probability of single-quasiparticle processes, producing the subsequent reduction 
of the probability of the Andreev processes, which in turn leads to both the 
suppression of the SGS and the reduction of the effective charge. 

\begin{figure}[t]
\begin{center}
\includegraphics[width=\columnwidth,clip=]{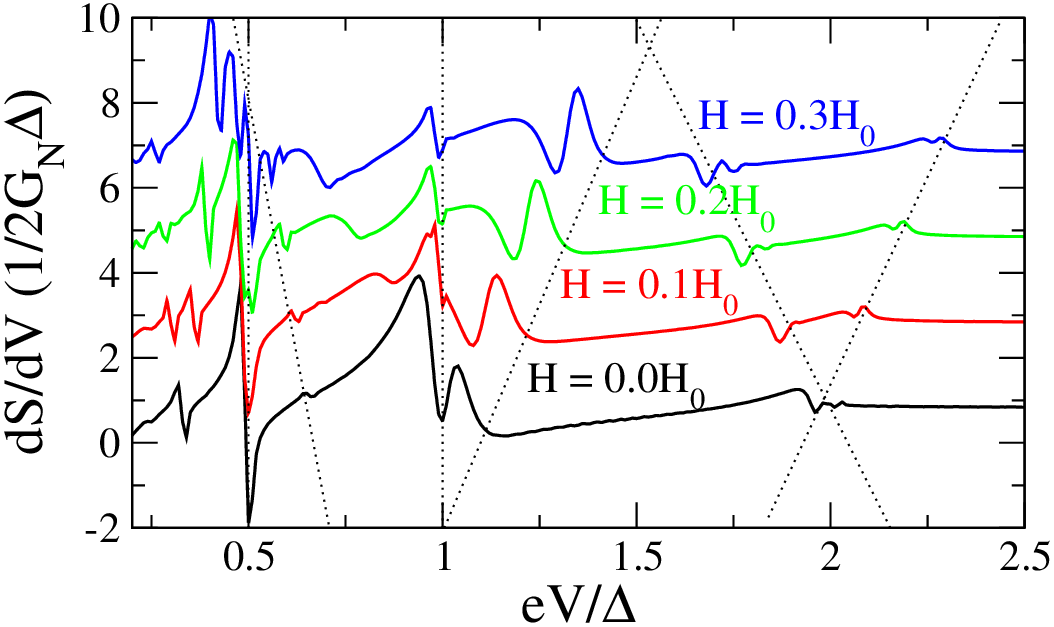}
\vspace*{-0.3truecm}
\caption{Differential shot noise of a clean $d_{\pi/4}$-$d_{-\pi/4}$ junction with 
$D=0.2$ and different values of the magnetic field. The curves has been vertically 
displaced for clarity and dotted lines have been added to guide the eye.}
\label{B-field}
\end{center}
\end{figure}
\vspace{-4mm}

Fogelstr\"om \emph{et al.}~\cite{Fogelstrom1997} have shown that the Andreev bound 
states should split in the presence of a magnetic field perpendicular to the $ab$-plane.
This splitting results in a splitting of the zero bias conductance peak observed in
tunnel junctions \cite{Aprili1999}. It is then interesting to analyze what is the 
signature of this time reversal symmetry breaking in the shot noise. Let us consider a 
magnetic field perpendicular to the $ab$-plane, $\vH=H\hat{\vz}$. As mentioned above,
this field leads to a Doppler shift in the continuum excitations given by 
$\vv_f\cdot\vp_s$, where the condensate momentum is $\vp_s = -(e/c)A(x)\hat{\vy}$, 
with $A$ the self-consistently determined vector potential \cite{Fogelstrom1997}. 
This means that the Andreev bound states are shifted to an energy which, in the limit 
of a large ratio $\lambda/\xi_0$, can be estimated to be $\epsilon_b(\pfhv) = (e/c)v_f 
H \lambda \sin \pfhv$, $\lambda$ being the $ab$-plane penetration depth. We shall use 
a natural field scale set by a screening current of order the bulk critical current, 
$H_0 = c\Delta/ev_f\lambda$, which is of the order of a Tesla \cite{Fogelstrom1997}. 
The screening currents flow parallel to the interface and in opposite directions in 
both electrodes, which means that the trajectory resolved DOS of the left and right 
superconductors are shifted by $2\epsilon_b(\pfhv)$ relative to each other. As 
explained in Ref.~\cite{Poenicke2002}, this shift modifies the threshold voltages 
of MARs starting and ending in different electrodes, leading to the splitting of 
the peaks with an odd order $n$ in the SGS. On the contrary, since the magnetic 
field produces a rigid shift of the spectrum, the threshold voltages of those 
MARs starting and ending in the same electrode are not modified. This means that 
the positions of the structures with an even order $n$ in the SGS remain unchanged. 
This is illustrated in Fig.~\ref{B-field} where we show the differential shot noise, 
$dS/dV$, for a $d_{\pi/4}$-$d_{-\pi/4}$ junction with transmission $D=0.2$ for different
values of the magnetic field. Starting at large voltages, the structure at $2\Delta$ 
splits with applied field. Around $eV=\Delta$ there is a maximum at $eV=\Delta$, 
unaffected by the applied field, as well as a field-shift of the peak just above 
$\Delta$. The field dependence of the differential noise is most clearly resolved 
at larger biases, $eV\ge \Delta/2$, as the marks of the various processes begin to 
overlap at small bias. The effect of the Doppler shift on the SGS of the noise is 
only prominent in junctions with a sizable misorientation. For junctions close to 
the $d_{0}$-$d_{0}$ case, the main contribution to the SGS comes from trajectories 
close to perpendicular incidence, i.e. with $\sin \pfhv \sim 0$ and thus having a 
vanishing Doppler shift.

In summary, we have presented a theory of shot noise in d-wave/d-wave contacts with
arbitrary transparency. We have shown that in the MAR regime these nonequilibrium
current fluctuations exhibit very peculiar features like subharmonic gap structure,
super-Poissonian noise ($S \gg 2eI$), reduction of the effective charge $q$ by 
impurities and the splitting of the SGS in a magnetic field. All these features
are unique fingerprints of the d-wave scenario, and we hope that our analysis will 
trigger off experimental study of shot noise in cuprate junctions.

JCC acknowledges the financial support provided by the EU under contract 
HPRN-CT-2000-00144, Nanoscale Dynamics, and MF by the Swedish research council. 

\vspace{-3mm}


\begin{thebibliography}{}
\bibitem{Blanter2000}
Ya.M.~Blanter and M.~B\"uttiker, Phys. Rep. {\bf 336}, 1 (2000).
\bibitem{Cron2001}
R.~Cron \emph{et al.}, Phys. Rev. Lett. {\bf 86}, 4104 (2001).
\bibitem{Cuevas1999}
J.C.~Cuevas, A.~Mart\'{\i}n-Rodero and A.~Levy~Yeyati, Phys. Rev. Lett. 
{\bf 82}, 4086 (1999).
Y.~Naveh and D.V.~Averin, Phys. Rev. Lett. {\bf 82}, 4090 (1999).
\bibitem{Zhu1999}
J.X.~Zhu and C.S.~Ting, Phys. Rev. B {\bf 59}, 14165 (1999).
Y.~Tanaka \emph{et al.}, Phys. Rev. B {\bf 61}, 11902 (2000).
T.~L\"ofwander, V.S.~Shumeiko and G.~Wendin, Physica C {\bf 367}, 86 (2002).
\bibitem{Harlingen1995}
D.J.~Van Harlingen, Rev. Mod. Phys. {\bf 67}, 515 (1995).
\bibitem{Tsuei2000}
C.C.~Tsuei and J.R.~Kirtley, Rev. Mod. Phys. {\bf 72}, 969 (2000).
\bibitem{Covington1997}
M.~Covington \emph{et al.}, Phys. Rev. Lett. {\bf 79}, 277 (1997).
\bibitem{Neils2002}
W.~K.~Neils and D.~J.~Van~Harlingen, Phys. Rev. Lett. {\bf 88}, 047001 (2002).
\bibitem{Saminadayar1997}
L.~Saminadayar \emph{et al.}, Phys. Rev. Lett. {\bf 79}, 2526 (1997).
R.~de-Picciotto \emph{et al.}, Nature {\bf 389}, 162 (1997).
\bibitem{Kashiwaya2000}
For a recent review on tunneling effect in unconventional superconductors see
S.~Kashiwaya and Y.~Tanaka, Rep. Prog. Phys. {\bf 63}, 1641 (2000).
\bibitem{Alff1998}
L.~Alff \emph{et al.}, Phys. Rev. B {\bf 58}, 11197 (1998).
L.~Alff \emph{et al.}, Eur. Phys. J. B {\bf 5}, 423 (1998).
\bibitem{Cuevas2001}
J.C.~Cuevas and M.~Fogelstr\"om, Phys. Rev. B {\bf 64}, 104502 (2001).
\bibitem{Matsumoto1995}
M.~Matsumoto and H.~Shiba, J. Phys. Soc. Jpn {\bf 64}, 3384 (1995); 
{\bf 64}, 4867 (1995).
\bibitem{Buchholtz1995}
L.J.~Buchholtz \emph{et al.}, J. Low Temp. Phys.  {\bf 101}, 1079 (1995); 
{\bf 101}, 1099 (1995).
\bibitem{Fogelstrom1997}
M.~Fogelstr\"om, D.~Rainer, and J.A.~Sauls, Phys. Rev. Lett. {\bf 79}, 281 (1997).
\bibitem{Barash1997}
Yu.S.~Barash, A.A.~Svidzinsky, and H.~Burkhardt, Phys. Rev B {\bf 55}, 15282 (1997).
\bibitem{Poenicke1999}
A.~Poenicke \emph{et al.}, Phys. Rev. B {\bf 59}, 7102 (1999).
\bibitem{Poenicke2000}
A.~Poenicke, M.~Fogelstr\"om, and J.A.~Sauls, Physica B {\bf 284-288}, 537 (2000).
\bibitem{Lofwander2001}
T. L\"ofwander, V.S. Shumeiko, and G. Wendin, Supercond. Sci. Technol {\bf 14} R53 (2001)
and references therein.
\bibitem{Poenicke2002}
A.~Poenicke, J.C.~Cuevas and M.~Fogelstr\"om, cond-mat/0203517, to appear in Phys. Rev. B
(2002).
\bibitem{Hu1994}
C.R.~Hu, Phys. Rev. Lett. {\bf 72}, 1526 (1994).
\bibitem{Constantinian2002}
K.Y.~Constantinian \emph{et al.}, Physica C {\bf 367}, 276 (2002).
\bibitem{Dieleman1997}
P.~Dieleman \emph{et al.}, Phys. Rev. Lett. {\bf 79}, 3486 (1997).
\bibitem{Aprili1999}
M.~Aprili, E.~Badica, and L.H.~Greene, Phys. Rev. Lett. {\bf 83}, 4630 (1999);
R.~Krupke and G.~Deutscher, Phys. Rev. Lett. {\bf 83}, 4634 (1999).
\end{thebibliography}
\end{document}